# Measurement of the neutron lifetime using an asymmetric magneto-gravitational trap and *in situ* detection

**Authors:**
R. W. Pattie Jr.[1], N. B. Callahan[2], C. Cude-Woods[1,3], E. R. Adamek[2], L. J. Broussard[4], S. M. Clayton[1], S. A. Currie[1], E. B. Dees[3], X. Ding[5], E. M. Engel[6], D. E. Fellers[1], W. Fox[2], P. Geltenbort[7], K. P. Hickerson[8], M. A. Hoffbauer[1], A. T. Holley[9], A. Komives[10], C.-Y. Liu[2], S. W. T. MacDonald[1], M. Makela[1], C. L. Morris[1], J. D. Ortiz[1], J. Ramsey[1], D. J. Salvat[11], A. Saunders[1], S. J. Seestrom[1†], E. I. Sharapov[12], S. K. Sjue[1], Z. Tang[1], J. Vanderwerp[2], B. Vogelaar[5], P. L. Walstrom[1], Z. Wang[1], W. Wei[1], H. L. Weaver[1], J. W. Wexler[3], T. L. Womack[1], A. R. Young[3], and B. A. Zeck[1,3]

**Affiliations:**
[1] Los Alamos National Laboratory, Los Alamos, NM 87545, USA
[2] Center for Exploration of Energy and Matter and Department of Physics, Indiana University, Bloomington, IN 47408, USA
[3] Triangle Universities Nuclear Laboratory and North Carolina State University, Raleigh, NC 27695, USA
[4] Oak Ridge National Laboratory, Oak Ridge, TN 37831, USA
[5] Department of Physics, Virginia Polytechnic Institute and State University, Blacksburg, VA 24061, USA
[6] West Point Military Academy, West Point, NY 10996, USA
[7] Institut Laue-Langevin, Grenoble, France
[8] Kellogg Radiation Laboratory, California Institute of Technology, Pasadena, CA 91125, USA
[9] Department of Physics, Tennessee Technological University, Cookeville, TN 38505, USA
[10] Department of Physics and Astronomy, DePauw University, Greencastle, IN 46135-0037, USA
[11] Department of Physics, University of Washington, Seattle, WA 98195-1560, USA
[12] Joint Institute for Nuclear Research, Dubna, Moscow region, Russia, 141980

*Correspondence to: ASaunders@lanl.gov

† Current address: Sandia National Laboratories, Albuquerque, NM 87185

**Abstract**: The precise value of the mean neutron lifetime, $\tau_n$, plays an important role in nuclear and particle physics and cosmology. It is a key input for predicting the ratio of protons to helium atoms in the primordial universe and is used to search for new physics beyond the Standard Model of particle physics. There is a 3.9 standard deviation discrepancy between $\tau_n$ measured by counting the decay rate of free neutrons in a beam (887.7 ± 2.2 s) and by counting surviving ultracold neutrons stored for different storage times in a material trap (878.5±0.8 s). The experiment described here eliminates loss mechanisms present in previous trap experiments by levitating polarized ultracold neutrons above the surface of an asymmetric storage trap using a repulsive magnetic field gradient so that the stored neutrons do not interact with material trap walls and neutrons in quasi-stable orbits rapidly exit the trap. As a result of this approach and the use of a new *in situ* neutron detector, the lifetime reported here (877.7 ± 0.7 (stat) +0.4/-0.2 (sys) s) is the first modern measurement of $\tau_n$ that does not require corrections larger than the quoted uncertainties.

**One Sentence Summary:** The first modern measurement of the free neutron lifetime with corrections smaller than the uncertainty is presented (877.7±0.7 +0.4/-0.2 s)

**Main Text:**
**Introduction**

Measurement of free neutron decay to a proton, electron, and antineutrino, $n \rightarrow p + e^- + \overline{\nu}_e$, the simplest nuclear beta-decay process, provides information about the fundamental parameters of the charged weak current of the nucleon and constraints on many extensions to the Standard Model at and above the TeV scale. The mean lifetime of this decay, $\tau_n$, played a role in determining the helium to hydrogen ratio in Big Bang Nucleosynthesis (BBN) [1]. In combination with neutron beta-decay correlations, $\tau_n$ can also be used to constrain Standard Model parameters and thus to search for new physics, including tensor and scalar currents in the semileptonic charged-current Lagrangian[2]. However, knowledge of $\tau_n$ to an accuracy of better than 1 s is necessary to improve BBN predictions of elemental abundances [1] and to search for physics beyond the Standard Model of nuclear and particle physics. The neutron lifetime has recently been measured by two different techniques [3, 4]: counting the surviving ultracold neutrons after storage in material-walled traps, with a most precise result of 878.5 ± 0.8 s [5]; and counting the number of decay products emerging from a passing beam of cold neutrons, with a result of 887.7 ± 2.2 s [6]. The results of these techniques disagree by 9.2 s, or 3.9 standard deviations.

*This manuscript has been authored by UT-Battelle, LLC under Contract No. DE-AC05-00OR22725 with the U.S. Department of Energy. The United States Government retains and the publisher, by accepting the article for publication, acknowledges that the United States Government retains a non-exclusive, paid-up, irrevocable, worldwide license to publish or reproduce the published form of this manuscript, or allow others to do so, for United States Government purposes. The Department of Energy will provide public access to these results of federally sponsored research in accordance with the DOE Public Access Plan (http://energy.gov/downloads/doe-public-access-plan).*

The present experiment was designed to reduce systematic uncertainties by using ultracold neutrons (UCN) trapped in a storage volume closed by magnetic fields on the bottom and sides and by gravity on top, as previously demonstrated by Ezhov *et al* [7]. In this work, we have used an asymmetric trap to reduce the population of long-lived closed neutron orbits with kinetic energies over the storable energy threshold in the trap [8, 9]. We have also introduced for the first time *in situ* detection of the surviving neutrons, to eliminate uncertainties associated with transporting the neutrons to an *ex situ* detector. Recently published storage experiments used storage traps with variable volumes to extrapolate to infinite volume in an attempt to reduce uncertainties associated with losses of neutrons due to interactions with the material walls [5, 10-14]. The present experiment had no detectable losses of neutrons due to interactions with the magnetic and gravitational "walls" of the trap and thus required no extrapolation. In addition, the present measurement used a number of techniques to diagnose and eliminate potential effects due to *quasi-trapped* neutrons. These neutrons have kinetic energies above the trapping potential but nevertheless can reside in the trap in quasi-stable orbits for hundreds of seconds, skewing the long storage time measurements.

**Experimental technique**
The experimental technique was described in detail in Morris *et al.* [15] and is summarized here. The experimental apparatus is shown in Figure 1. UCN were supplied by the west beam line of the LANSCE UCN facility at Los Alamos National Laboratory [16, 17]. The UCN flux was monitored by normalization detectors [18] that sampled the UCN flux through small aperture holes in the beam guide. The neutrons were polarized by a 6 T solenoid magnet that transmitted only neutrons in the "high-field seeking" spin state. An adiabatic fast passage spin flipper changed the spin state of the neutrons to low-field seeking with ~90% efficiency, using an applied magnetic field of 14 mT and an oscillating radio-frequency field of 372 kHz. Another normalization detector mounted above the storage height of the neutron trap monitored the UCN flux just before the entrance to the trap. The neutrons reaching this detector height had too much kinetic energy to be confined by gravity in the magnetic trap. The flux at each detector was monitored throughout the filling period to determine the relative normalization of the number of neutrons entering the trap in each storage run. Neutrons filled the trap through a movable magnetic door located at the bottom of the apparatus.

The trap was constructed of a Halbach array of permanent magnets in which the magnetization of each row of permanent magnets was rotated 90º relative to its neighbors. Each NdFeB magnet was 2.54 cm x 5.08 cm x 1.27 cm, with a surface field of about 1.0 T. The magnets were installed along the surface of two intersecting tori, one with a major radius of 100 cm and a minor radius of 50 cm, the other with the radii interchanged, and cut off at a height of 50 cm from the bottom of the trap, thus forming an asymmetric trap with a trapping potential of about 50 neV (corresponding to neutron temperature ≲ 0.58 mK) and a fiducial volume of 420 l. An additional externally applied holding field of up to 10 mT, approximately perpendicular to the Halbach field, was used to maintain the neutron polarization during the storage period. The performance of the trap was described in Salvat *et al* [9].

At the end of the 150 s filling period, the 800 MeV proton beam that produced the UCN was turned off to reduce backgrounds, and the loading trap door and other valves in the UCN beam pipe were closed, preventing further neutrons from reaching the apparatus. A cleaning period followed, designed to eliminate any neutrons in the trap with kinetic energy sufficient to escape the trap. A horizontal sheet of polyethylene, called the "cleaner", removed neutrons with sufficient kinetic energy to reach its height via absorption or thermal upscattering. During the filling and cleaning periods, the cleaner was positioned 38 cm above the bottom of the trap, or 12 cm below the nominal open top of the trap. The cleaner covered approximately one half of the horizontal surface of the trap at its mounted height (~0.86 m$^2$, to be compared with the 0.23 m$^2$ horizontal cleaner used in the work of Ref[15]), so that every neutron capable of reaching it did so quickly within a few tens of seconds after entering the trap. A second "active" cleaner, with a 28% of the area, was mounted on the downstream side of the trap in the same plane as the primary cleaner. This second cleaner used $^{10}$B-coated ZnS:Ag as the UCN absorber [18] and was observed by an array of PMTs, allowing the UCN density in the plane of the cleaners to be continuously monitored. At the conclusion of the cleaning period, 50-300 s for the data presented here, both cleaners were raised 5 cm to stop further interactions; at this point the storage period began. The neutrons were stored for times typically ranging from 10 s to 1400 s, chosen to optimize statistical reach in a given experimental running time while still permitting systematic studies.

At the end of the storage period, a UCN detector consisting of a vertical poly(methyl methacrylate) (PMMA) paddle coated on both sides with ZnS:Ag and $^{10}$B, with a total active surface area of 750 cm$^2$ per side (~21% of the 3450 cm$^2$ area of the midplane of the trap), was lowered into the center of the trap. The detector could be lowered in multiple steps, and, at its lowest position, reached to within 1 cm of the bottom of the trap. This permitted rate-dependent uncertainties to be studied by controlling the counting rate and also enabled the exploration of different neutron energy- and phase space-dependent systematic effects. The detector removed the surviving stored neutrons from the trap with a time constant of about 8 s. At the conclusion of the counting period (typically 100 to 300 s in length, or many 8 s mean draining times), the detector was left in the trap to count background rates with no neutrons in the trap for typically 150 s. The absolute efficiency of the detector, previously reported to be 96%[15], and those of the upstream monitor detectors, cancel in the ratios used to measure the lifetime in this experiment. Each neutron absorbed on the detector's boron layer generated a burst of scintillation photons in the ZnS:Ag scintillator that were converted and conducted from the transparent PMMA backing plate to a pair of photomultiplier tubes by an array of 2 mm spaced wavelength shifting fibers. The photons in each PMT were individually counted with an 800 ps precision time stamp by an input channel of the same multichannel scaler (MCS) [19] that counted the output pulses from the normalization monitors.

In a typical measurement cycle, a pair of runs were performed, one with a nominal short storage time of 20 s and one with a nominal long storage time of 1020 s, each with ~2.5×10$^4$ neutrons in the trap at the beginning of the storage period. A total of 332 pairs of long and short runs were analyzed for the results in this paper, in five different running configurations. The different run conditions varied the cleaning time, the number of steps in which the detector was lowered into the trap, and the magnitude of the applied neutron polarization holding field. The five run conditions are listed in Table 1.

Figure 2 shows a nine-step (left side) and three-step (right side) unloading curve, for a short and long storage time, summed over all the cycles in the respective run condition. In each case, the first, highest detector step placed the bottom edge of the detector at the cleaning height so that no stored neutrons had sufficient energy to reach the 132 cm$^2$ of active area that extended below the position of the raised cleaners. No neutrons above background were detected in this step, putting constraints on systematic uncertainties caused by insufficient cleaning of high energy neutrons and heating of neutrons by vibrations, to be described later: only eight peaks are visible in the "nine-step" unloading curve and two in the "three-step" curve. The absolute time of the long storage curve was adjusted in each plot by the difference between the nominal long and short storage times, to allow visual comparison of the curves.

The data were blinded by adjusting the nominal storage times (along with all the MCS timestamps in a blinded run) by a factor hidden from those analyzing the data, with the result that $\tau_n$ extracted from the blinded data differed from the actually measured value by a random offset of up to ±15 s.

**Analysis**
The expected number of surviving neutrons after storing an initial number $N_0$ of neutrons in the trap for a time *t* is $N_{surv}=N_0 e^{-t/\tau_{meas}}$ where $\tau_{meas}$ is the mean measured survival time of the trapped neutrons. The measured loss rate is the sum of the loss rate due to neutron decay and all other sources of loss from the trap, such as losses caused by interactions with the walls, depolarization of neutrons during storage, thermal upscattering of neutrons from residual gas in the trap, or other sources of loss: $1/\tau_{meas}=1/\tau_n+1/\tau_{loss}$.

The number of surviving neutrons was estimated from the raw data, consisting of a string of time-stamped photon events from the two *in situ* detector PMTs, using two different techniques. The first method required a coincidence between photons from each of the two PMTs to identify a neutron. Coincidences were identified during a 50 ns window, followed by an above threshold number of photons in a variable integration window of several microsecond length. The threshold was determined by the number and arrival time of previously identified neutrons in the data stream. Identification of new neutrons was disabled during the integration window, creating a rate-dependent but calibrated software dead time for each counting bin. The integration window was extended in 1 μs steps as long as photon events continued to arrive, to maximize neutron identification efficiency while minimizing software dead time. The second method used individual photons, or "singles" data. Each individual photon was treated as an independent event, with no attempt made to identify the neutron responsible for each individual photon.

The normalized total signals of the surviving UCN populations, or yields, were calculated for each run by summing the counts measured in all detector positions, subtracting experimental backgrounds, and dividing by the relative number of neutrons loaded into the trap. The background at long holding times was of order 0.4% of signal for coincidence counting and 15% for singles counting. The raw numbers of neutrons (coincidence) or photons (singles) were corrected for dead time. The relative number of neutrons loaded into the trap was determined from the counts in the elevated normalization monitor (see Fig. 1), exponentially weighted by the measured loading time constant of the trap (60-70 s). Spectral variations in the incident neutron flux were assessed by taking the ratio of the number of counts in the elevated normalization detector to the number in the upstream, beam-height normalization detector. The first order correction to the normalization, as much as 10% over a period of 100 hours, was determined by minimizing the variance of neutron yields of short-storage runs only. Alternating long and short storage time runs reduced the effect of this correction on the lifetime uncertainty to negligible levels.

The neutron lifetime and uncertainty were calculated from pairs of short and long storage time yields using:

$$R_i \equiv \frac{Y_{is}}{Y_{il}}$$

$$\tau_{n,i} = \frac{(\overline{t_l} - \overline{t_s})}{\ln(R_i)} \text{ and} \qquad\qquad 1)$$

$$\Delta\tau_{n,i} = \frac{(\overline{t_l} - \overline{t_s})}{\ln(R_i)^2} \frac{\Delta R_i}{R_i}$$

where the subscripts $l$ and $s$ denote long and short storage times, $\overline{t}$ is the mean neutron detection time during the counting period, Y's are the yields, $\tau_n$ is the lifetime, the $\Delta$'s indicate uncertainties, and the subscript $i$ denotes the individual run pairs. Uncertainties were calculated using Poisson statistics for the UCN yields including the statistics of the exponentially-weighted elevated normalization monitor counts. The statistical uncertainty in the photon singles yields were obtained from the coincidence data and by analyzing the variance in the singles-extracted lifetimes, and both approaches produced consistent results.

Average lifetimes were calculated for each of the five run configurations in three ways. First, the average long and short yields across a run configuration were calculated and assigned uncertainties as the standard deviation of the individual yields divided by $\sqrt{N}$, where $N$ is the number of individual run pair yields. The lifetimes and uncertainties were calculated from the average values using Eq. 1. The second method determined the lifetime from a weighted average of the lifetimes calculated from each individual run pair. In this case, the uncertainty was calculated from the weighted average of individual run pair uncertainties and multiplied by the square root of the reduced chi square to account for any remaining non-statistical variation in the data set caused by smaller systematic effects such as higher order time or spectral variations in the loaded neutrons. The third method was identical to the second, except that an unweighted average was used to compute the average lifetime.

**Systematic Uncertainties**
The total systematic uncertainty in these results were estimated to be 0.28 s. The major sources of systematic uncertainties in these results are listed in Table 2.

The only correction applied to the central value of the lifetime was due to thermal upscattering of UCN from individual interactions of UCN with residual gas particles in the trap during the storage period. This pressure correction was made using two calibrated cold cathode gauges located above the mid-plane of the trap to measure the pressure (typically $6 \times 10^{-7}$ Torr), a residual gas analyzer to measure the mass spectrum of the residual gas, and measured UCN cross sections [20, 21] to calculate the velocity-independent UCN lifetime due to losses on the residual gas in the trap. The loss rate from this lifetime was then subtracted from the measured neutron loss rate to yield the neutron decay loss rate, for a correction to the measured lifetime of 0.16 s with an uncertainty of 0.03 s.

The systematic uncertainty due to possible depolarization of neutrons during the storage period was assessed by measuring the neutron lifetime while varying the magnitude of the applied polarization holding field. High precision lifetime measurements were made at holding field strengths of 6.8 mT and 3.4 mT, and lower precision measurements at smaller fields down to 0.5 mT. The resulting lifetimes were fitted using a power law suggested by calculations of depolarization in the present trap geometry by Steyerl et al. [22]:

$$\frac{1}{\tau} = \frac{1}{\tau_n} + \frac{B_{\perp 0}^2}{B_\perp^2 \tau_{DP}},$$

where $B_\perp$ is the holding field, $B_{\perp 0}$ is the full holding field, $\tau$ is the measured storage lifetime, $\tau_n$ is the neutron decay lifetime, and $\tau_{DP}$ is the loss lifetime due to depolarization. The result of the fit yielded a loss lifetime due to depolarization of $\tau_{DP} = 1.1 \times 10^7$ s (with one sigma uncertainty bounds of $6.0 \times 10^6$ s and $5.5 \times 10^7$ s) for an uncertainty on the measured neutron lifetime of 0.07 s. The 6.8 mT and 3.4 mT measurements showed no variation outside of statistics.

Neutrons can be heated by, for example, many small interactions with the vibrational motion of the UCN trap's magnetic field, slowly gaining enough energy to exceed the trap potential and escape from the trap during the long storage period. A limit on the uncertainty due to this effect was determined by looking for neutrons moving into the highest neutron detector position (38 cm above the bottom of the trap, or equal in height to the lowered cleaner), in run configurations B – E during the long storage time (see Table 1).

The detector's ability to count these very low energy (<5 neV kinetic energy at the top of their orbits in the trap) UCN was verified by loading the trap with the cleaners and the neutron detector in their raised position. The counting curve from the active cleaner was observed to fall to its background rate in ~5 s and was constant thereafter, indicating that neutrons were efficiently cleaned to the height of the raised cleaners (43 cm above the bottom of the trap). The neutron detector was then lowered to the upper counting position and was observed to count UCN with sufficient energy to reach heights between 38 cm and 43 cm with a time constant of 260 s. While only the lowest 5 cm of the detector was exposed below the raised cleaners during the first step of a nominal lifetime run, characterization of the dagger detector with a Gd-148 alpha particle source showed an approximately uniform optical response across the entire active area of the detector. In addition, heated UCN gain energy slowly and so have multiple opportunities to be captured by the detector before acquiring sufficient energy to reach the position of the raised cleaners. In estimating the cleaning and heating systematic uncertainties, a conservative factor of twenty was applied to account for neutrons that were not detected during the 20 s first-position counting time used for lifetime running.

The number of neutrons observed in the highest counting position was consistent with background in the long storage time runs. The systematic uncertainty due to heating was determined by the uncertainty of the yield calculated using only the counts observed in the highest position. Based on this analysis, we put a 1-σ limit on the lifetime uncertainty due to neutron heating of 0.23 s.

Similarly, insufficient cleaning of neutrons with energy above the trapping energy would result in an artificial excess of counts after the short storage time that could be lost over the long storage time. An analysis of excess neutrons in the highest counting position in the short storage time runs, also consistent with zero neutrons above background, allowed us to put a limit of 0.06 s on the uncertainty in the measured lifetime due to insufficient cleaning.

In the singles photon counting analysis method, a hardware dead time was caused by the 10 ns dead time of the discriminator used to detect the individual photons. The uncertainty due to correcting for this rate-dependent effect was set to 20% of the correction on each run, for a total uncertainty on the extracted lifetime of 0.04 s.

Phase space evolution can cause a possible change in effective detector efficiency between the short and long storage time runs due to evolution of the neutron population in the trap between regions of phase space with different degrees of access to the detector location. Any effect of phase space evolution on the measured lifetime would cause a variation in the relative number of neutrons in the peaks corresponding to the different counting steps between the short and long storage time runs; therefore, a limit on this effect was estimated by calculating the statistical uncertainty of the centroid of the unloading curve for the nine-step measurements, and was found to be 0.10 s. The effect was smaller for the one and three step measurements. A cross check was made for the presence of neutron population phase space evolution between the short and long storage periods by comparing the number of counts in each of the three or nine separate detector steps; the variation in these ratios was consistent with statistical fluctuations.

**Results**

As a final check for non-exponential behavior in the data, a global fit to the yields of the long and short storage time measurements was performed and is shown in Figure 3. The unblinded measured neutron lifetime extracted from this fit was 877.6 ± 0.7 s with a $\chi^2$/dof of 0.7. After correction for gas upscattering, the final unblinded measured mean neutron lifetime was 877.7 ± 0.7 (stat) +0.4/-0.2 (sys) s. The coincidence and the singles analysis methods described above yielded the same results. Three independent analyses were conducted and compared before unblinding. These analyses agreed to within 0.2 s. The central value of the result presented here is the average of the three results, the statistical uncertainty is the average of those from the three analyses, and the systematic uncertainty is that from Table 2 added in quadrature with an additional 0.2 s to account for the differences between the analysis techniques. Because the total uncertainty of this result is dominated by statistical uncertainty, and because the leading systematic uncertainties appear to be statistically-driven and thus reducible with further study, we expect to ultimately reach a total uncertainty well below 0.5 s in future data runs using this apparatus.

The non-blinded data set presented in Ref [15] was combined with a blinded systematics study data set, which had a statistical accuracy of $\Delta\tau_n$ = 1 s, to develop techniques to correct for incomplete cleaning of quasi-bound neutrons and to identify improvements to the trap cleaning procedure. These improvements were implemented prior to acquiring the data discussed in this paper. The systematics study data were unblinded (based on two independent analyses) at the same time as the data presented here and produced a consistent result for $\tau_n$. This data set was not included in the neutron lifetime result presented here.

The result presented here is the first modern measurement of the neutron lifetime that does not require corrections to the measured lifetime that are larger than the quoted uncertainties. This result agrees with the previous best measurement of the lifetime for neutron decay using UCN stored in a material trap, and disagrees with the lifetime for neutron beta-decay determined using the beam technique.

# References and Notes:

1. Cyburt, R.H., *et al*., Big bang nucleosynthesis: Present status. Reviews of Modern Physics, 2016. 88(1): p. 015004.
2. Marciano, W.J., Fundamental neutron physics. Physics Procedia, 2014. 51: p. 19.
3. Wietfeldt, F. E. and Greene, G. L. Colloquium: the neutron lifetime. Reviews of Modern Physics, 2011. 83: p. 1173.
4. Young, A. R. *et al*., Beta decay measurements with ultracold neutrons: a review of recent measurements and the research program at Los Alamos National Laboratory. Journal of Physics G: Nuclear and Particle Physics, 2014. 41: p. 114007.
5. Serebrov, A., *et al*., Neutron lifetime measurements using gravitationally trapped ultracold neutrons. Physical Review C, 2008. 78(3): p. 035505.
6. Yue, A., *et al*., Improved determination of the neutron lifetime. Physical Review Letters, 2013. 111(22): p. 222501.
7. Ezhov, V., *et al*., Magnetic storage of UCN for a measurement of the neutron lifetime. Nuclear Instruments and Methods in Physics Research Section A: Accelerators, Spectrometers, Detectors and Associated Equipment, 2009. 611(2): p. 167; Ezhov, V. et al., Measurement of the neutron lifetime with ultra-cold neutrons stored in a magneto-gravitational trap. arXiv preprint arXiv:1412.7434, 2014.

8. Walstrom, P., *et al*., A magneto-gravitational trap for absolute measurement of the ultra-cold neutron lifetime. Nuclear Instruments and Methods in Physics Research Section A: Accelerators, Spectrometers, Detectors and Associated Equipment, 2009. 599(1): p. 82.
9. Salvat, D. J. *et al*., Storage of ultracold neutrons in the magneto-gravitational trap of the UCN τ experiment. Physical Review C, 2014. 89(5): p. 052501.
10. Arzumanov, S. *et al*., A measurement of the neutron lifetime using the method of storage of ultracold neutrons and detection of inelastically up-scattered neutrons.  Physics Letters B, 2015. 745: p. 79.
11. Steyerl, A. *et al*., Quasielastic scattering in the interaction of ultracold neutrons with a liquid wall and application in a reanalysis of the Mambo I neutron-lifetime experiment. Physical Review C, 2012. 85: p. 065503.
12. Pichlmaier, A. *et al*., Neutron lifetime measurement with the UCN trap-in-trap MAMBO II. Physics Letters B, 2010. 693: p. 221.
13. Byrne, J. *et al*., A revised value for the neutron lifetime measured using a Penning trap. Europhysics Letters 1996. 33: p. 187.
14. Mampe, W. *et al*., Measuring neutron lifetime by storing ultracold neutrons and detecting inelastically scattered neutrons. JETP Letters, 1993. 57: p. 82.
15. Morris, C., *et al*., A new method for measuring the neutron lifetime using an in situ neutron detector. Review of Scientific Instruments, 2017.  88(5): p. 053508.
16. Saunders, A., *et al*., Demonstration of a solid deuterium source of ultra-cold neutrons. Physics Letters B, 2004. 593(1): p. 55.
17. Saunders, A. *et al*., Performance of the Los Alamos National Laboratory spallation-driven solid-deuterium ultra-cold neutron source. Rev. Sci. Inst. 84, 013304 (2013).
18. Wang, Z., *et al*., A multilayer surface detector for ultracold neutrons. Nuclear Instruments and Methods in Physics Research Section A: Accelerators, Spectrometers, Detectors and Associated Equipment, 2015. 798: p. 30.
19. FastCom. 2016  [cited 2016; Available from: https://www.fastcomtec.com/fwww/datashee/photon/mcs6.pdf].
20. Seestrom, S. J., *et al*., Upscattering of ultracold neutrons from gases. Physical Review C, 2015. 92(6): p. 065501.
21. Seestrom, S. J., *et al*., Total cross sections for ultracold neutrons scattered from gases. Physical Review C, 2017. 95(1): p. 015501.
22. Steyerl, A., *et al*., Spin flip loss in magnetic confinement of ultracold neutrons for neutron lifetime experiments. Physical Review C, 2017. 95(3): p. 035502.

**Acknowledgments:**

This work was supported by the Los Alamos Laboratory Directed Research and Development (LDRD) office (No. 20140568DR), the LDRD Program of Oak Ridge National Laboratory, managed by UT-Battelle, LLC (No. 8215), the National Science Foundation (Nos. 130692, 1307426, 161454, 1306997, and 1553861), NIST Precision Measurement Grant, IU Center for Space Time Symmetries (IUCSS), and DOE Low Energy Nuclear Physics (Nos. DE-FG02-97ER41042 and DE-AC05-00OR22725). The authors would like to thank the staff of LANSCE for their diligent efforts to develop the diagnostics and new techniques required to provide the proton beam for this experiment.

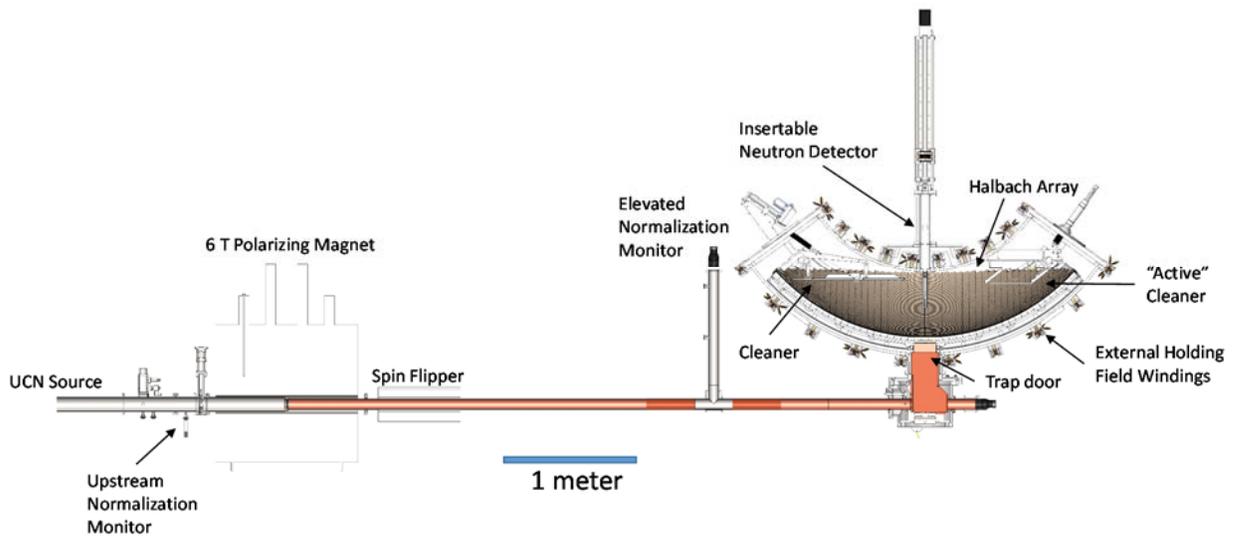

**Fig. 1**. Layout of the UCN beam line and trap used for these measurements

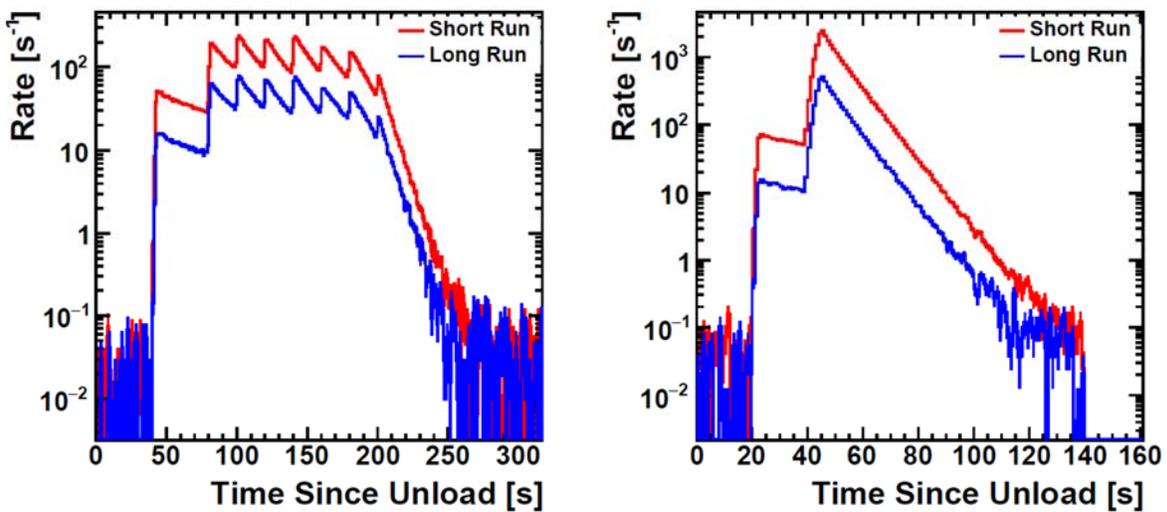

**Fig. 2.** Left) Nine step unloading time distributions. Right) Three step unloading distributions. The times have been shifted to align the short (red) and long (blue) distributions. These distributions include both the neutron signals and backgrounds.

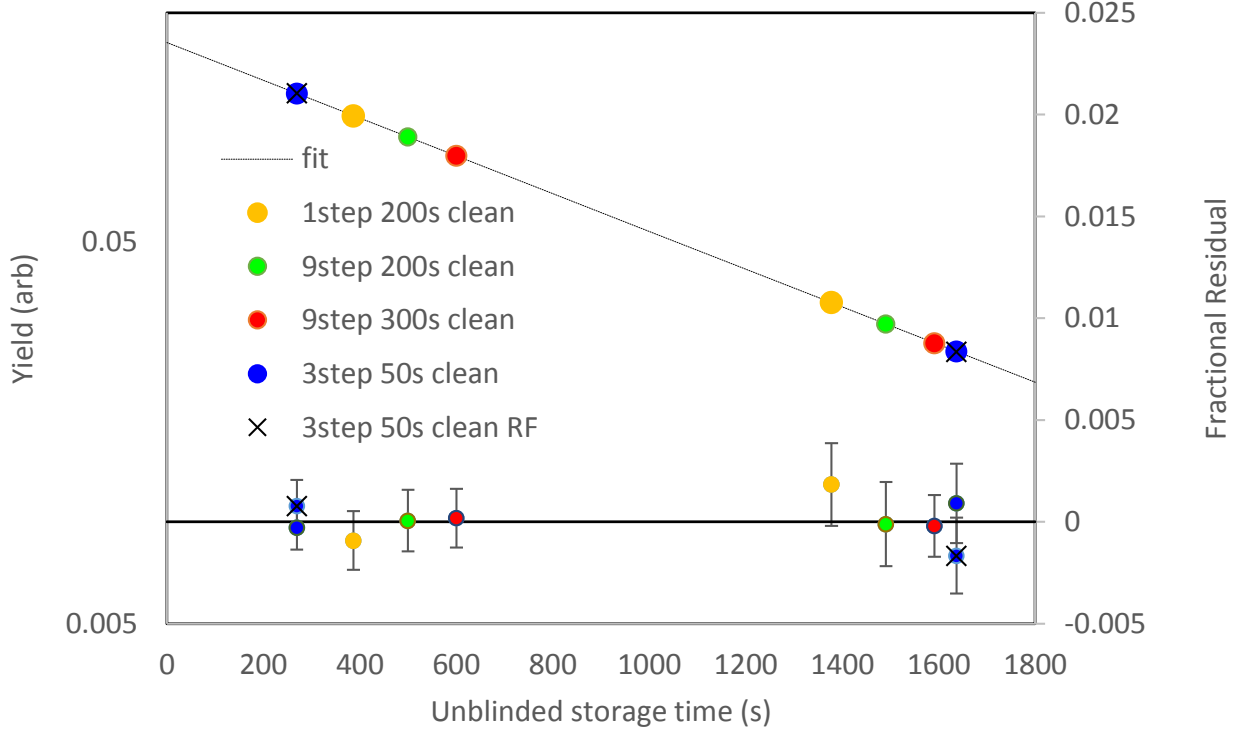

**Fig. 3.** Plot of the yields as a function of time relative to the start of filling for each of the running conditions. The relative normalization of the data sets has been adjusted to account for the different running conditions. The fractional residuals are plotted on the bottom axis. "RF" denotes the use of a reduce depolarization holding field strength of 3.4 mT instead of 6.8 mT.

**Table 1.** A summary of the five running conditions analyzed in this paper. "Detector Steps" is the number of discrete counting positions as the neutron detector was lowered to the bottom of the trap; "Cleaning Time" is the length of cleaning period from the closing of the neutron loading trap door to the raising of the cleaner; "Holding Field" is the minimum strength of the externally applied polarization holding field in the trap; and "Run Pairs" is the number of long-short run pairs acquired for this configuration, all with roughly equal numbers of initially loaded neutrons.

| Run configuration | Detector Steps | Cleaning Time (s) | Holding Field (mT) | Run Pairs |
|---|---|---|---|---|
| A | 1 | 200 | 6.8 | 79 |
| B | 9 | 200 | 6.8 | 66 |
| C | 9 | 300 | 6.8 | 70 |
| D | 3 | 50 | 6.8 | 60 |
| E | 3 | 50 | 3.4 | 57 |

**Table 2.** Systematic uncertainties.

| Effect | Upper bound (s) | Direction | Method of evaluation |
|---|---|---|---|
| Depolarization | 0.07 | + | Varied external holding field |
| Microphonic heating | 0.24 | + | Detector for heated neutrons |
| Insufficient cleaning | 0.07 | + | Detector for uncleaned neutrons |
| Dead time/pileup | 0.04 | ± | Known hardware dead time |
| Phase space evolution | 0.10 | ± | Measured neutron arrival time |
| Residual gas interactions | 0.03 | ± | Measured gas cross sections and pressure |
| Background shifts | <0.01 | ± | Measured background as function of detector position |
| Total | 0.28 | | (uncorrelated sum) |